\documentstyle[preprint,aps]{revtex}
\begin{document}
\title{Confronting particle emission scenarios with strangeness data}
\author{Fr\'ed\'erique Grassi$^{1,2}$ and Otavio Socolowski Jr.$^3$}
\address{$^1$Instituto de F\'{\i}sica, Universidade de S\~{a}o Paulo, \\
C. P. 66318, 05315-970 S\~{a}o Paulo-SP, Brazil\\
$^2$Instituto de F\'{\i}sica, Universidade Estadual de Campinas, \\
C. P. 1170, 13083-970 Campinas-SP, Brazil\\
$^3$Instituto de F\'{\i}sica Te\'{o}rica, UNESP, Rua Pamplona 145, \\
01405-901 S\~{a}o Paulo-SP, Brazil}
\maketitle

\begin{abstract}
\noindent 
We  show that a hadron gas model with continuous particle emission instead
of freeze out may solve some of the problems
(high values of the freeze out density and specific net charge)
 that one encounters
in the latter case
when studying strange particle ratios such as those by WA85.
This underlines the necessity to understand better particle emission in 
hydrodynamics to be able to analyze data. It also 
re-opens the possibility of
a quark-hadron transition occuring with phase equilibrium instead of 
explosively.
\end{abstract}

\bigskip\ 

{\bf 1. Introduction}
 
An enhancement of strangeness production in relativistic nuclear collisions
(compared to e.g. proton-proton collisions at the same energy) is a possible
signature \cite{ko86}
of the much sought-after quark-gluon plasma.
It is therefore particularly interesting that current data at AGS
(Alternating Gradient Synchroton) and SPS (Super Proton Synchrotron)
energies do show an increase in strangeness production
(see e.g. \cite{qm96}). At SPS energies, this increase seems to imply 
that something new is happening: in microscopical models, one has to
postulate some previously unseen reaction mechanism (color rope formation
in the RQMD code
\cite{so92}, multiquark clusters in the Venus  code \cite{ai93},
etc) while hydrodynamical models have their own problems
(be it those with a rapidly hadronizing plasma \cite{le95a}
or those with an equilibrated hadronic phase, preceded or not
by a plasma phase). 
In this paper, we examine
the
shortcomings of the latter class of hydrodynamical models and suggest that they
might be due to
a too rough description of particle emission\footnote{The main problem for 
the former class of hydrodynamical models is the difficulty to yield enough
entropy after hadronization.}. 

To be more precise, let us assume that a hadronic fireball 
(region filled with a hadron gas, or HG,
 in local thermal and chemical equilibrium)
is formed
in heavy ion collisions
at SPS energies and that particles are emitted from it at freeze out
(i.e. when they stop interacting due to matter dilution).
One then runs into (at least) three kinds of problems when 
discussing strange particle ratios.

First, the temperature ($T_{f.out} \sim$ 200 MeV) and
baryonic potential ($\mu_{b\,f.out} \sim$ few 100 MeV) needed {\em at freeze out} 
\cite{da91a,da91b,vo92,da92,cl93a} 
 to reproduce WA85 \cite{wa85new}
and NA35 \cite{na35}
strangeness
data actually correspond to {\em high}
 particle densities:
this is inconsistent with the very notion of freeze out
\footnote{While WA85 and NA35 data for strange particle ratios are comparable
and lead to high T's and $\mu_b$'s, NA36 data are different and lead to lower
T's (Phys. Lett. B327 (1994) 433) for similar targets but a somewhat different 
kinematic window.  However the rapidity distribution for $\Lambda$'s (E.G. Judd
et al. Nucl. Phys. A590 (1995) 291c) as well as $\overline{\Lambda}$'s and
$K^0_s$'s (J.Eschke et al. Heavy Ion Phys.4 (1996) 105) for NA36 are quite below
that of NA35; NA44 midrapidity data for $K^{\pm}$ agree with that of NA35.}.

Second, 
to reproduce strange particle ratios, it turns out that
the strange quark potential $\mu_s$
must be small and the strangeness saturation
factor  $\gamma_s$ of order 1
(this quantity, with value usually between 0 and 1,
 measures how far from chemical equilibrium the
strange particles are,  1 corresponds to full
chemical equilibrium of the strange particles). 
Both facts are expected
in a quark-gluon plasma  hadronizing suddenly,
not normally in a hadronic fireball \cite{le92,so94}.

Third, 
using the values at freeze out
 of the temperature, baryonic potential
and saturation factor extracted
 to reproduce WA85 strange particle ratios,
one can predict the value of another quantity, the
 specific net charge  (ratio of the net charge multiplicity
to the total charge multiplicity). This quantity has been measured not by WA85,
but in 
experimental conditions similar to that of WA85 by EMU05
\cite{emu05}. 
It turns out that the predicted value is too high, 
(while it might be smaller if a quark-gluon plasma 
fireball had been formed)
\cite{le93,le95a}.

In what follows, we study
how problems 1 and 3 are related to the mechanism for particle emission
normally used, freeze out, and suggest that the use of 
continuous emission instead of freeze out, might
 shed some light on these questions.
(We also re-discuss problem 2.)
This underlines
the necessity to understand better particle emission in 
hydrodynamics and re-opens perspectives (see conclusion)
 for scenarios of the quark-hadron 
transition.

\bigskip\

{\bf 2. Fluid behavior and particle spectra}

First let us see in more details what are the two particle emission mechanisms
just mentioned.
In the usual freeze-out scenario, hadrons are kept in thermal
equilibrium until some decoupling criterion has become satisfied
(then they free-stream toward the detectors). 
For example, in the papers mentioned above where experimental
strange particle ratios
are reproduced, the
freeze out criterion is that a certain
 temperature and baryonic potential have been reached.
The formula for the emitted particle spectra used normally is the Cooper-Frye
formula \cite{co74}. 
In the particular case of a gas expanding longitudinally only
in a boost invariant way, freezing out at some fixed
temperature and chemical potential, the Cooper-Frye formula reads
\begin{equation}
\frac{dN}{dy p_{\perp} dp_{\perp}} =\frac{gR^2}{2 \pi} \tau_{f.out} m_{\perp} 
\sum_{n=1}^{\infty} (\mp)^{n+1} \exp{(n\mu_{f.out}/T_{f.out})}
 K_1(n m_{\perp}/T_{f.out}).                
\end{equation}
(The plus sign corresponds to bosons and minus, to fermions.)
It depends only on the conditions at freeze out: $T_{f.out}$ and 
$\mu_{f.out}=\mu_{b\,f.out}B+\mu_{S\,f.out}S$, with B and S the baryon number
and strangeness of the hadron species considered, and 
$\mu_{S\,f.out}(\mu_{b\,f.out},T_{f.out})$ obtained by imposing strangeness
neutrality. 
So the experimental spectra of particles teach us in that case
only what the conditions were at freeze out.
 
In the continuous
emission scenario developed in \cite{gr95a,gr95b}, the basic idea is
the following. 
Due to the finite dimensions 
and lifetime of the fluid, 
 a particle at  space-time point $x$,
has some chance $\cal P$ to have already made its last collision.
In that case, it will fly freely towards the detector, carrying with it memory of
what the conditions were in the fluid at $x$.
Therefore the spectrum of emitted particles contains an integral over all 
space and time,
counting particles where they last interacted. In other words, the
experimental spectra will give us in principle information about
the whole fluid history, not just the freeze out conditions.
For the case of a fluid expanding longitudinally only
in a boost invariant way with continuous particle emission, 
the formula that parallels (1) is
\begin{equation}
\frac{dN}{dy p_{\perp} dp_{\perp}} \sim
\frac{2 g}{(2 \pi)^2} 
\int_{{\cal P}=0.5} d\phi  d \eta   
\frac{m_{\perp} \cosh \eta \tau_F \rho d\rho       
+ p_{\perp} \cos \phi \rho_F \tau d\tau }
{\exp((m_{\perp}\cosh \eta-\mu) /T) \pm 1} 
\end{equation}
where $\tau_F(\rho,\phi,\eta;v_{\perp}$ 
(resp. $\rho_F(\tau,\phi,\eta;v_{\perp})$) is the time (resp. radius) where
the probability to escape without collision
${\cal P}=0.5$ is reached.
${\cal P }$ is given by a Glauber formula, 
$\exp [-\int \sigma v_{rel} n(\tau') d\tau']$ and depends in particular
on location
and direction of motion.
We are using both (1) and (2) in the following.
Clearly, 
in (2), various $T$ and $\mu=\mu_b B + \mu_S S$ appear
(again $\mu_S(\mu_b,T)$ is obtained from strangeness neutrality), 
reflecting the whole fluid history,
not just $T_{f.out}$ and $\mu_{b\,f.out}$.

So to predict particle spectra, in the case of continuous emission,
we need to know the fluid history. To get it, we fix some initial conditions
$T(\tau_0,\rho)=T_0$ and $\mu_b(\tau_0,\rho)=\mu_{b\,0}$ and
solve the equations of conservation of 
momentum-energy and baryon number for a mixture of free and interacting 
particles, using the equation of
 state of a resonance gas (including the 207 known lowest mass particles)
and imposing strangeness neutrality.
As a result we get $T(\tau,\rho),\mu_b(\tau,\rho)$ and we can use these as input 
in the formula for the particle spectra (2). The procedure is similar to
that of a massless pion gas \cite{gr95a,gr95b}
 but is numerically more involved.

An important result 
\cite{gr95a,gr95b}
for the following is that for heavy particles, the
spectrum (2) is dominated by the initial conditions, precisely a formula similar
to (1) with freeze out quantities replaced by initial conditions could
be used as an approximation, (particularly at high $p_{\perp}$);
for light particles the whole fluid history
matters. To understand this fact, one can consider 
equation 2  and compare particles
emitted at $T(\tau,\rho)$=200 and 100 MeV. 
For particles with mass 
of 1 GeV, the exponential term gives a thermal suppression above one hundred
between these two tempratures.
The multiplicative factors in front of the exponential are
in principle larger at the lower temperature but do not compensate for
such a big decrease. This is why heavy particles are abundantly
emitted at high temperatures.
On the other side for pions,
the thermal suppression is only a factor of 2. This is why  light particles
are emitted significantly in a larger interval of temperatures. 

Note that since heavy particle and high $p_{\perp}$ particle spectra
are sensitive mostly to the initial values of T and $\mu_b$, the exact fluid expansion 
does not matter very much for them; in particular the assumption of boost 
invariance should play no part in the forthcoming analysis of strange
high $p_{\perp}$ particle ratios\footnote{Note also that the data considered 
below are in a small rapidity window, near midrapidty. Were it not for this
fact, boost invariance should not be assumed, because the rapidity 
distributions do not have this symmetry.}. 
It would be however interesting to
include continuous emission in, e.g. a hydrodynamical code, to obtain the
fluid evolution and study pion data and low $p_{\perp}$ data.

\bigskip\

{\bf 3. Particle ratios}

Once the spectra have been obtained, they can be integrated to get particle
numbers,
taking into account eventual experimental cutoffs
and correcting  for resonance decays. Since we had to specify  
the initial conditions to solve the conservation equations
 and use this solution as input into (2), the
particle numbers depend on $T_0,\mu_{b\,0}$.
In contrast, for the freeze out case, particle numbers
depend on the conditions at freeze out,
$T_{f.out}$ and $\mu_{b\,f.out}$.

We look for regions in the $T_o,\mu_{b\,0}$ space which permit to reproduce the
latest WA85 experimental data on strange baryons \cite{wa85new}
for $2.3<y<2.8$ and $1.0<p_{\perp}<3.0$ GeV:
$\bar{\Lambda}/\Lambda=0.20\pm 0.01$,
$\overline{\Xi^-}/\Xi^-=0.41 \pm 0.05$ and
$\Xi^-/\Lambda=0.09 \pm 0.01$
($\overline{\Xi^-}/\bar{\Lambda}=0.20 \pm 0.03$
follows). In fact there is no set of
initial conditions which permits to reproduce all the above ratios. 
A similar
situation occurs with freeze out, as noted in \cite{re93}.
(Note that with the older experimental data \cite{wa85old}, a region in the 
$T_{f\,out}-\mu_{b\,f\,out}$ space
permiting to reproduce all the above ratios 
existed \cite{da91a,vo92,da92,cl93a}.)

In the comparison of our model with WA85 data we have assumed however
complete  chemical equilibrium for strangeness production. As already 
mentioned in the introduction, this is not expected for a HG. In order to 
account for incomplete strangeness equilibration, we introduce
the additional strangeness saturation parameter $\gamma_s$
by making the substitution $\exp {\mu_S S} \rightarrow \gamma_s^{\mid S \mid}
\exp {\mu_S S}$ in the (Boltzmann) distribution functions
\cite{ra91}.
In our case, a priori,
$\gamma_s$ depends on the space-time location $x$, 
however since as already mentioned,
the initial conditions dominate in the shape and normalization of the spectra
of heavy particles (particularly at high $m_{\perp}$), 
we take
\begin{equation}
\frac{dN}{dy p_{\perp} dp_{\perp}} \sim
\gamma_s^{\mid S \mid}(\tau_0) \frac{dN_{eq}}{dy p_{\perp} dp_{\perp}} 
\end{equation}
with $dN_{eq}/dy p_{\perp} dp_{\perp}$ given by (2). In figure 1a, we
see that for $\gamma_s(\tau_0)=0.58$, there exists an overlap region in
the 
$T_0,\mu_{b\,0}$ plane where all the above ratios are reproduced.
For the freeze out case, a similar situation occurs as noted in \cite{re93},
namely there exists an overlap region for $\gamma_s=0.7$.
 
In the freeze out case,
the values of the parameters in the overlap region  correspond
to high particle densities 
and so it is hard to understand how
particles have ceased to interact: this is the problem 1 mentioned
in the introduction.
In the continuous emission case, $T_0$ and $\mu_{b\,0}$ 
in the overlap region lead to high initial densities,
but this is
of course quite reasonable since these are values when the HG 
started its hydrodynamical expansion. Note also that the value we used for
$< \sigma v_{rel}>$ comes from using a Breit-Wigner formula for
the $\Lambda-\pi$ cross section and computing 
$< \sigma v_{rel}>_{p_{\perp}^{\Lambda} \geq 1.0}$ at various temperatures.

The aim of picture 1a is to allow an easy comparison with freeze out
results such as \cite{re93},
 however it is not physically complete: so far we have 
neglected hadronic volume corrections. 
For freeze out, this correction cancels between numerator and denominator in
baryon ratios so can be ignored \cite{cl93a}
but for continuous emission, since we are considering the 
whole fluid history to get particle numbers (and then their ratio),
it must be included.
There are various ways to do this.
(Some of the)
thermodynamical quantities for pointless particles can be multiplied by a
factor (determined from pointlike quantities), for example 
$1/(1+V n_0)$, where $V$ is a typical hadronic volume
and 0 indicates pointlike quantities. 
However there is no consensus on whether this correction involves a
net baryonic density $n_0$ 
\cite{cl86,cl90,cl93a,cl92}
 or the total particle density 
$n_0=\sum_j n^0_j$
\cite{ri91,cl93b}
or a compromise 
\cite{vo92}. Here
we modify all particle densities  as follows $n_i= 
n_i^0 /(1+V \sum_j n^0_j)$ as in \cite{ri91,cl93b}.
Taking into account hadronic volumes, we get the overlap region
shown in figure 1b, which is  shifted towards smaller T's and 
$\mu_b$'s but not very different from that of figure 1a. 
Given that simulations
of QCD on a lattice indicate a quark-hadron transition for 
temperatures\footnote{In these simulations, the transition temperature depends
strongly on the number of flavors $N_f$ (being $>$ 200 MeV for $N_f=0$ and
$<$ 200 MeV for $N_f>1$) and somehow on the method of
extraction. Since the created plasma might remain gluon dominated
with quarks away from equilibrium,
up to the transition
(this is already predicted to be the case  for RHIC and LHC energies), it is
not clear what value of the transition temperature should be used
but one should probably not take results from lattice QCD as a
rigid value in our context.} around
200 MeV,
it seems more reasonable to consider initial conditions
$T_0 \sim$ 190 MeV and $\mu_{b\,0} \sim$ 180 MeV, i.e. the bottom part of the overlap
region. The {\em precise} location of the overlap region (and exact value of 
$\gamma_s$)
is sensitive
to  changes in the equation of state -as we have just seen- as well as
 in the cross section or cutoff ${\cal P}=0.5$ in eq. 2.
(As a cross check, we have also
remade our calculations using a Walecka-type equation of state 
\cite{me93} and found similar results to figure 1b.)
Therefore problem 1 (whether the overlap region is physically reasonable)
is taken care of.

To be complete, we also examined the more recent ratios
obtained by WA85 
\cite{wa85prel} (at midrapidity):
$\overline{\Omega^-}/\Omega^-_{m_{\perp} \geq 2.3 GeV}=0.57\pm 0.41$,
$(\Omega^-+\overline{\Omega^-})/(\Xi^-+\overline{\Xi^-})_{m_{\perp} \geq 2.3 
GeV}= 1.7 \pm 0.9$, 
$K^0_s/\Lambda_{p_{\perp} > 1.0 GeV}=1.43\pm 0.10$,
$K^0_s/\bar{\Lambda}_{p_{\perp} > 1.0 GeV}=6.45\pm 0.61$ and
$K^+/K^-_{p_{\perp} > 0.9 GeV}=1.67\pm 0.15$.
We looked for a region in the
$T_0,\mu_{b\,0}$ plane where 
$\overline{\Omega^-}/\Omega^-_{m_{\perp} \geq 2.3 GeV}$ is reproduced: due
to the large experimental error bars, this does not bring new restrictions
to fig. 1b.
We also calculated our value for
$(\Omega^-+\overline{\Omega^-})/(\Xi^-+\overline{\Xi^-})_{m_{\perp} \geq 2.3 
GeV}$ in
the overlapping region and found $\sim 0.7$, in marginal agreement with the 
above experimental values.
The three ratios involving
kaons depend on more than just initial conditions (kaons are intermediate
in mass between pions and lambdas so part of the fluid thermal history 
must be refleted in their spectra), in particular $\gamma_s(x) \sim  
\gamma_s(\tau_0) \sim cst$ may
 not be a good approximation and we are still working on this.
The above experimental
ratios concern SW collisions, data with SS are not so extensive 
yet  
but not very different 
\cite{wa85ss}
so a similar overlapping region can be found.

The apparent temperature extracted from the 
experimental $p_{\perp}$ spectra for $\Lambda$,
$\overline{\Lambda}$, $\Xi^-$ and $\overline{\Xi^-}$ is
$\sim$ 230 MeV \cite{wa85new}. Given that we extracted from ratios of
these particles, temperatures T$_0$ $\geq$ 200 MeV, we conclude that heavy 
particles exhibit little transverse flow,
which is compatible with the fact 
that they are emitted early during the hydrodynamical 
expansion\footnote{Note that WA85 data concern 
high $p_{\perp}$ strange particles. But in \cite{gr95a,gr95b}, we showed that
(a simpler version of) our model with continuous emission and no tranverse
flow, reproduces the shape of the NA35 S+S
$p_{\perp}$ spectra for strange particles,
which extend to low $p_{\perp}$ (but are not restricted to midrapitidy).}.

\bigskip\

\noindent{\bf 4. Specific net charge}

We now turn to 
\begin{equation}
D_q=(N^+-N^-)/(N^++N^-)
\end{equation}
using the continuous emission scenario.
As mentioned in the introduction, for HG models
with freeze out the predicted $D_q$ is too high,
when using values of the freeze out parameters that fit strangeness data,
e.g. $T_{f.out}\sim$ 200 MeV, $\mu_{b\,f.out} \sim$ 200 MeV and $\gamma_s \sim $
0.7. 
For continuous emission, 
due to thermal suppression, particles
{\em heavier than the pion} are approximately emitted at  
$T_0 \sim $ 200 MeV, $\mu_{b\,0} \sim $ 200 MeV and $\gamma_s \sim $ 0.49
(fig. 1b),
so $D_q$ so far is similar to that of freeze out.
However there is an additional source of particles that enters the denominator
of (4),
namely pions are emitted at $T_0$ {\em and then on} (since they are not
thermally suppressed).
So we expect to get a lower value for $D_q$
 in the continuous emission case than in the freeze out case. 
(We recall that pions are the dominant contribution in $N^++N^-$.)
This would
go into the
direction of solving problem 3, it is
still under investigation.

\bigskip\

{\bf 5. Conclusion}

Our present
description is simplified, for example we do not include the transverse
expansion of the fluid, use similar interaction cross sections for all types
of particles, etc. In addition, we need to look systematically
at strangeness data from other collaborations as well as other types of data
such as Bose-Einstein correlations. Nevertheless,
we have seen that the continuous emission scenario with a HG may
shed light on
problems 1 and 3 (discussed in the introduction)
that a freeze out model with a HG encounters. Namely, 
in the overlap region of the parameters needed to reproduce WA85 data,
the density of particles is high and this is consistent with the emisssion
mechanism, since it is the initial density of the
thermalized fluid. We also expect
$D_q$ to be smaller for continuous emission than freeze out.
But (problem 2) the value 
of the strangeness saturation parameter may be  high for a HG, particularly
at the begining of its hydrodynamical expansion.
However what we really need to get fig. 1b, is that 
$\Xi^-/\Lambda = \gamma_{\Xi} \Xi^-/\gamma_{\Lambda} \Lambda_{\mid eq.}$
and 
$\overline{\Xi^-}/\overline{\Lambda} = \gamma_{\Xi} \overline{\Xi^-}
/\gamma_{\Lambda} \overline{\Lambda}_{\mid eq.}$ with $\gamma_{\Xi}/\gamma_{\Lambda}
=0.49$. We expect indeed that multistrange $\Xi^-$ and $\overline{\Xi^-}$
are more far off chemical equilibrium than singlestrange $\Lambda$ and
$\overline{\Lambda}$ so that $\gamma_{\Xi}/\gamma_{\Lambda} < 1$.
The result $\gamma_s=0.49$ arises {\em if} one makes the
{\em additional hypothesis} that quarks are independent degrees of freedom 
inside the hadrons so that one has factorizations of the type
$\gamma_{\Lambda} \exp {\mu_{\Lambda}/T} = \gamma_s \exp {2 \mu_q/T}
\exp {\mu_s/T}$ and 
$\gamma_{\Xi} \exp {\mu_{\Xi}/T} = \gamma_s^2 \exp { \mu_q/T}
\exp {2 \mu_s/T}$. Therefore problem 2 may not be so serious.

The fact that we may cure some of the problems of the HG scenario does not 
mean that no quark-gluon plasma has been created before the HG, in fact it may
open new possibilities for scenarios of
the quark-hadron transition
(e.g. an equilibrated quark-gluon plasma evolving into an 
equilibrated HG with continouous emission); in particular it may not be 
necessary to assume an explosive transition \cite{le95a}
or a deflagration/detonation scenario \cite{bi93,bi94,cs94}.

But our main conclusion is that the emission mechanism may modify profoundly
our interpretation of data
(for example, does the slope in transverse mass spectrum tell something
about freeze out or initial conditions?).
In turn this modifies our discussion of what potential problems 
(such as 1 and 3) are emerging.
Therefore we believe it is necessary to devote more work to get a realistic
description of particle emission in hydrodynamics, \cite{gr95a,gr95b} being
a first step in that direction.
We remind that
the idea that particles are emitted continuously and not on a sharp freeze 
out surface is supported by microscopical simulations at AGS energies 
\cite{br95} and SPS energies \cite{so96}.

\bigskip\
 
\noindent{\bf Acknowledgement}

The authors wish to thank U.Ornik for providing some of the computer programs
to start working on this problem.
This work was partially supported by FAPESP (proc. 95/4635-0),
CNPq
(proc. 300054/92-0) and CAPES.

\bigskip\
 
\noindent{\bf Note added}

After completing this paper, we learned that G.D. Yen, M.I. Gorenstein,
W. Greiner and S.N. Yang suggested \cite{ye97} another solution to problems
1 and 3 above, in term of the excluded volume approach of \cite{ri91}, for the
preliminary Au+Au (AGS) and Pb+Pb (SPS) data.

\newpage\ 

\begin{center}
{\bf FIGURE\ CAPTIONS}
\end{center}

\begin{description}
\item[FIG.1]  Overlap region in the $T_0-\mu_{b\,0}$ plane for
WA85 data, with $< \sigma v_{rel} > =1\, fm^2$
a) without 
b) with hadronic volume corrections. 
\end{description}


\begin{references}
\bibitem{emu05} To be published, quoted in [16] and [5].
\end{references}
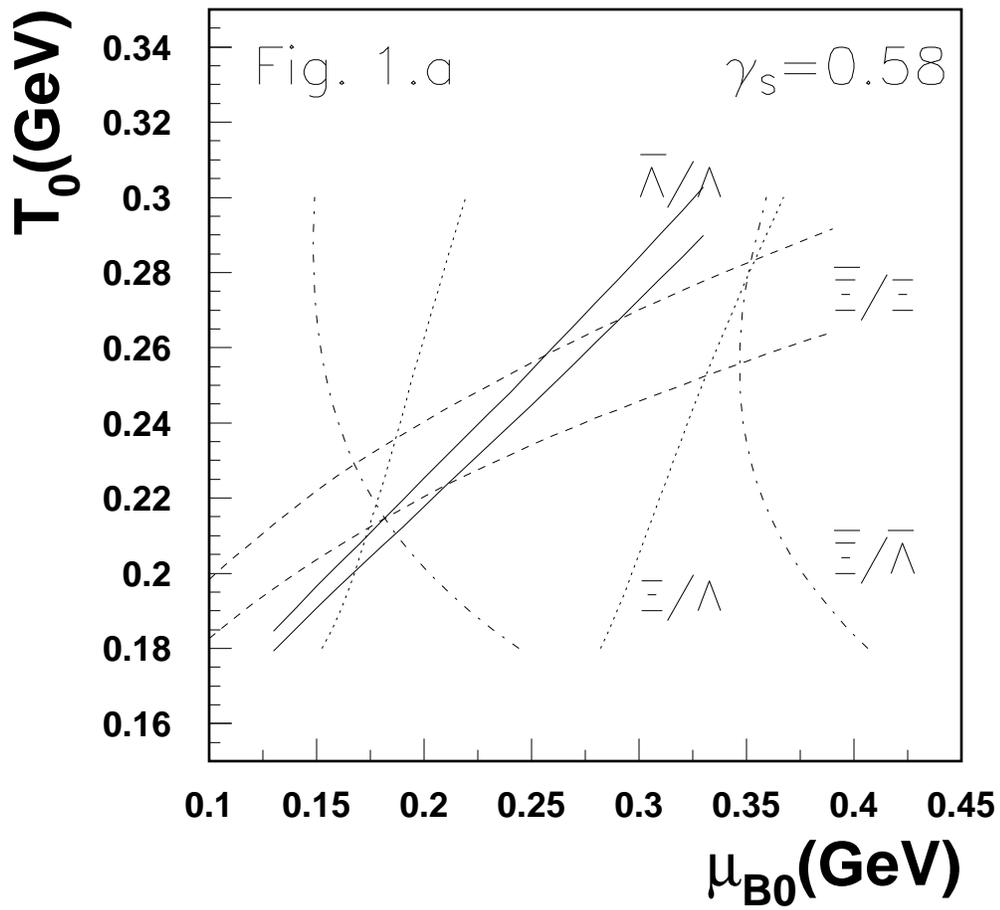
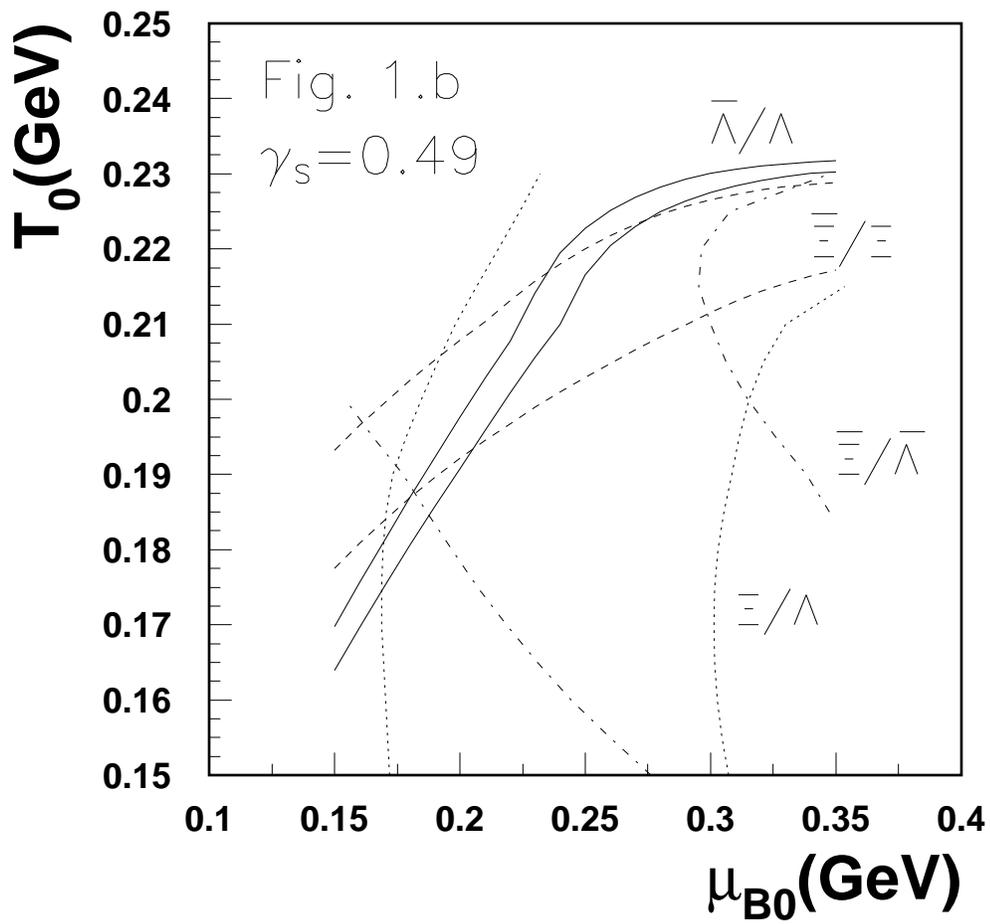
\end{document}